\input harvmac
\input epsf
\def\half{{\textstyle{1\over2}}}
\def\ga{\gamma}
\def\de{\delta}
\def\al{\alpha}
\def\ep{\epsilon}

\def\frak#1#2{{\textstyle{{#1}\over{#2}}}}

\def\sic{supersymmetric}

\def\pa{\partial}

\def\Dbar{\bar D}
\def \in{\leftskip = 40 pt\rightskip = 40pt}
\def \out{\leftskip = 0 pt\rightskip = 0pt}

\def\npb{{Nucl.\ Phys.\ }{\bf B}}

\def\plb{{Phys.\ Lett.\ }{\bf B}}

\def\sjnp{Sov.\ J.\ Nucl.\ Phys.\ }

\def\lf{16\pi^2}
\def\llf{(16\pi^2)^2}
\def\lllf{(16\pi^2)^3}
\def\llllf{(16\pi^2)^4}

{\nopagenumbers
\line{\hfil LTH 381 }
\line{\hfil hep-ph/9609325}
\vskip .5in
\centerline{\titlefont Scheme dependence and the NSVZ $\beta$-function.}
\vskip 1in
\centerline{\bf I.~Jack, D.R.T.~Jones and C.G.~North}
\bigskip
\centerline{\it DAMTP, University of Liverpool, Liverpool L69 3BX, U.K.}
\vskip .3in
We investigate the connection between  the NSVZ and the DRED forms 
of the gauge $\beta$-function in an $N=1$ supersymmetric gauge theory. 
We construct a coupling constant redefinition that 
relates the two forms up to four loops. By abelian calculations, we 
are able to infer the complete non-abelian form of $\beta_g^{(3)DRED}$, 
and also $\beta_g^{(4)DRED}$ except for one undetermined parameter. 

\Date{September 1996}}

\newsec{Introduction}
An all-orders formula for the gauge $\beta$-function in an $N=1$ supersymmetric
gauge theory was presented some years ago. This result (which we shall call
$\beta_g^{NSVZ}$) originally appeared (for the special case of 
no chiral superfields) in Ref.~\ref\tim{D.R.T.~Jones, \plb 
123 (1983) 45}, and was subsequently generalised, using instanton calculus,  
in Ref.~\ref\nov{
V.~Novikov et al, \npb 229 (1983) 381\semi
V.~Novikov et al, \plb166 (1986) 329\semi 
M.~Shifman and A.~Vainstein, \npb 277 (1986) 456}. 
(See also Ref.~\ref\shifa{A.~Vainstein, V.~Zakharov and M.~Shifman,
\sjnp 43 (1986) 1028\semi
M.~Shifman, A.~Vainstein and V.~Zakharov \plb 166 (1986) 334}.)
For a recent discussion emphasising the importance of holomorphy,   
see \ref\shif{M.~Shifman, hep-ph/9606281}.

Recently the renormalisation group fixed points of  
$\beta_g^{NSVZ}$ have become 
important in the study of duality (for a review, see Ref.~\ref\dual{
K.~Intriligator and N.~Seiberg, Nucl.\ Phys.\ Proc.\ Suppl.\ 45{\bf B/C} (1996)
1}). 
An interesting question, therefore, is as follows: given
the renormalisation scheme dependence of $\beta$-functions beyond one loop,
in which scheme is the NSVZ  result
valid? For instance, will calculations using standard 
dimensional reduction (DRED) give the NSVZ result? The DRED result certainly 
agrees with the NSVZ result at one and two loops; moreover certain properties
of $\beta_g^{DRED}$ at higher loops are consistent with the
NSVZ result. Namely, $\beta_g^{DRED}$ is known to vanish at three 
loops for a one-loop finite theory
\ref\pwb{A.J.~Parkes and P.C.~West, \npb256 (1985) 340}
\ref\gmz{M.T.~Grisaru, B.~Milewski and D.~Zanon, \plb 155 (1985)357}\
 and furthermore, if we specialise to the
$N=2$ case, $\beta_g$ and the anomalous dimensions of the
chiral superfields vanish beyond one loop\ref\hsw{
P.S.~Howe, K.S.~Stelle and P.~West, \plb 124 (1983) 55\semi
P.S.~Howe, K.S.~Stelle and P.K.~Townsend, \npb  236 (1984) 125}.
\foot{We will discuss later whether this result will apply in  schemes 
other than DRED.}
One might accordingly be tempted to speculate that DRED will 
reproduce the NSVZ formula to all orders. However, in a recent note
\ref\jjnb{I.~Jack, D.R.T.~Jones and C.G.~North, hep-ph/9606323}\ we 
showed that this is not the case; at three loops the DRED result is related to
the NSVZ result by a coupling constant redefinition. In the present 
paper we shall give more details of this calculation and also extend the
result to four loops, at least in the abelian case. 

Before proceeding, however, it is worthwhile emphasising the following
point.  It is sometimes asserted that the perturbative coefficients of
$\beta_g$ are  quite arbitrary beyond two loops, so that, for example,
all contributions  at three and more loops can be transformed to zero.
We shall see, however, that in the general case (with a superpotential)
the nature of  possible changes in $\beta_g$ 
due to  redefinitions $\delta g$ which are
manifestly gauge-invariant analytic functions of $g$ and the Yukawa
couplings $Y^{ijk}$  is
heavily constrained.  
(One has also the freedom  to make redefinitions
$\delta Y^{ijk}$, but, as we shall see, these are not germane to the
issue of whether the NSVZ and DRED results are equivalent.)

\newsec{The three-loop calculation}
The Lagrangian $L_{\rm SUSY} (W)$ for an $N=1$ \sic\ theory
is defined by the superpotential
\eqn\Ea{
W=\frak{1}{6}Y^{ijk}\Phi_i\Phi_j\Phi_k+\frak{1}{2}\mu^{ij}\Phi_i\Phi_j. }
$L_{\rm SUSY}$ is the Lagrangian for  the $N=1$ supersymmetric
gauge theory, containing the gauge multiplet $V$  
 and a multiplet of chiral superfields $\Phi_i$ with component fields
$\{\phi_i,\psi_i\}$, transforming as a
representation $R$ of the gauge group $\cal G$.
We assume that there are no gauge-singlet
fields.
The $\beta$-functions for the Yukawa couplings $\beta_Y^{ijk}$
are given by
\eqn\Ec{
\beta_Y^{ijk}= Y^{p(ij}\ga^{k)}{}_p =
Y^{ijp}\ga^k{}_p+(k\leftrightarrow i)+(k\leftrightarrow j),}
where $\ga$ is the anomalous dimension for $\Phi$.
The one-loop results for the gauge coupling $\beta$-function $\beta_g$ and
for $\ga$ are given by
\eqn\Ed{
\lf\beta_g^{(1)}=g^3Q,\quad\hbox{and}\quad
\lf\ga^{(1)i}{}_j=P^i{}_j,}
where
\eqna\Ee$$\eqalignno{
Q&=T(R)-3C(G),\quad\hbox{and}\quad &\Ee a\cr
P^i{}_j&=\frak{1}{2}Y^{ikl}Y_{jkl}-2g^2C(R)^i{}_j. &\Ee b\cr}$$
Here $Y_{jkl} = (Y^{jkl})^*$, and 
\eqn\Ef{
T(R)\delta_{AB} = \tr(R_A R_B),\quad C(G)\delta_{AB} = f_{ACD}f_{BCD}
\quad\hbox{and}\quad C(R)^i{}_j = (R_A R_A)^i{}_j.}
The two-loop
$\beta$-functions for the dimensionless couplings were calculated in
\break Refs.~\pwb, \ref\tja{D.R.T.~Jones, \npb87 (1975) 127}%
\nref\pwa{A.J.~Parkes and P.C.~West, \plb138 (1984) 99}%
\nref\west{P.~West, \plb137 (1984) 371}%
--\ref\tjlm{D.R.T. Jones and L. Mezincescu, \plb136 (1984) 242;
{\it ibid} 138 (1984) 293}:
\eqna\Au$$\eqalignno{ \llf\beta_g^{(2)}&=2g^5C(G)Q-2g^3r^{-1}C(R)^i{}_jP^j{}_i
&\Au a\cr
\llf\ga^{(2)i}{}_j&=[-Y_{jmn}Y^{mpi}-2g^2C(R)^p{}_j\delta^i{}_n]P^n{}_p+
2g^4C(R)^i{}_jQ,&\Au b\cr}
$$
where $Q$ and $P^i{}_j$ are given by Eq.~\Ee{}, and $r=\delta_{AA}$.

In our notation the NSVZ formula for $\beta_g$ is
\eqn\russa{\beta_g^{NSVZ} =
{{g^3}\over{\lf}}\left[ {{Q- 2r^{-1}\tr\left[\ga C(R)\right]}  
\over{1- 2C(G)g^2{(\lf)}^{-1}}}\right],}
which leads to
\eqn\russab{\eqalign{\lllf\beta^{(3)NSVZ}_g &=
4g^7 Q C(G)^2 -4g^5 C(G) r^{-1} \lf \tr\left[ \ga^{(1)}C (R)\right]
\cr&-2g^3 r^{-1} \llf \tr\left[ \ga^{(2)}C (R)\right].\cr}}
As mentioned in the Introduction, one would like to know in which 
renormalisation scheme this result is valid, since $\beta$-functions are 
scheme dependent. It is easily seen from Eqs.~\Ed\ and \Au{}\ that the NSVZ
result coincides with the results of DRED up to two loops. Will DRED reproduce
the NSVZ result to all orders? There are two pieces of evidence
which appear to favour this conjecture. Firstly, $\beta_g^{(3)DRED}$ 
has been explicitly shown to vanish for a 
one-loop finite theory\pwb, i.e. one for which $P=Q=0$, and it is clear from 
Eqs.~\russab\ and \Au{}\ that this property is shared by $\beta_g^{(3)NSVZ}$. 
A second piece of evidence comes from specialising to the
$N=2$ case. 
In $N = 1$ language, an $N =2$ theory is defined by the superpotential
\eqn\agb{
W = \sqrt{2}g \eta_A \chi^i S_A{}^j{}_i \xi_j
}
where $\eta, \chi$ and $\xi$ transform according to the adjoint, $S^*$
and $S$  representations respectively. The set of chiral superfields
$\chi, \xi$  is called a hypermultiplet. 
$N=2$ theories have one-loop divergences only\hsw;
using DRED
we may therefore expect that $\beta_g$ and the anomalous 
dimension of both
the $\eta$ and the hypermultiplet should vanish beyond one loop.
\foot{Clearly the absence of divergences beyond one loop  implies that 
$\beta_g$ vanishes beyond one loop as long as  
minimal subtraction is employed;  higher order contributions to the $\beta_g$ 
can be  invoked by making finite subtractions, or equivalently by 
a redefinition $g\to g + \delta g$.}
We see that the NSVZ result of Eq.~\russa\ is consistent with this property;
we have
\eqn\agc{\eqalign{
P_{\eta AB} &= Qg^2\de_{AB}\cr
P_{\chi} &= P_{\xi} = 0,\cr}}
so that if $\ga$ vanishes beyond one loop then Eq.~\russa\ reduces to
$\beta_g^{NSVZ} ={{g^3}\over{\lf}}Q$, which is of course the one-loop result.
In particular, since from Eq.~\Au{b}\ we have $\gamma^{(2)}=0$ for $N=2$, we 
have $\beta_g^{(3)NSVZ}=0$ for an $N=2$ theory.  
Nevertheless, despite these indications that the NSVZ result might coincide
exactly with that obtained using DRED, we shall now show that 
in fact the NSVZ and DRED results part company at three loops. We shall see that
they are related by a coupling constant redefinition corresponding to a change 
of renormalisation scheme. We shall calculate $\beta_g^{(3)DRED}$ 
explicitly in the abelian case and construct the coupling constant 
redefinition which effects the transition to the NSVZ result. We shall then 
extend this redefinition to the full non-abelian case by exploiting the 
known $N=2$ properties of the DRED result, and hence we shall deduce 
the full non-abelian $\beta_g^{(3)DRED}$. 

We calculate $\beta_g$ by computing the divergences in the 
vector field two-point function, using the super-Feynman gauge.  This is
sufficient since in the abelian case the background superfield
calculation and the normal superfield calculation are identical. The
relevant diagrams are shown in Fig.~1. The shaded blobs represent one-loop
self-energy insertions. We use dimensional reduction
(DRED) with minimal subtraction, setting $\epsilon = 4 - d$. 
The divergent part of each individual diagram,
after performing  all the appropriate subtractions for divergent
sub-diagrams, will be expressible as a  combination
$A\Pi_{1\over2}+B\Pi_0$, where $A$ and $B$ are analytic in 
$1\over{\ep}$ and the coupling constants, and 
$\Pi_{1\over2}$ and $\Pi_0$ are projection operators defined in the 
Appendix.
with $\Pi_{1\over2}+\Pi_0=-\partial^2$. Upon adding the results for all
diagrams, the total will only involve $\Pi_{1\over2}$, reflecting the
transversality of the vector propagator. We use a convenient and
efficient short-cut to calculate the total coefficient for
$\Pi_{1\over2}$ without calculating the full contribution for each
diagram. The idea is the following: To obtain the total coefficient of
$\Pi_{1\over2}$, it will be  sufficient to know the difference $B-A$ 
for each diagram. Upon summing over all diagrams, the sum over
the $B$s will give zero just leaving the sum over the $A$s, which is
what we want. The point is that we can obtain the combination $2(B-A)$
simply by adding up the divergent contribution to $D^2\Dbar^2$ from each
 diagram, regarding the derivatives as exactly anticommuting. (So that 
$D^2\Dbar^2$, $\Dbar^2D^2$ and $D^{\al}\Dbar^2D_{\al}$ would each count
the same.) Each diagram may start with up to 10 $D$s and 10 $\Dbar$s 
(all on internal lines). For each diagram, we manipulate the
supercovariant derivatives $D$ and $\Dbar$ using integration by parts
and the anticommutation relations Eq.~(A.1) 
until we obtain a set of diagrams each
of which contains 8 $D$s and 8 $\Dbar$s, possibly together with some
ordinary derivatives. During this process we avoid  integrating any $D$
or $\Dbar$ onto an external line. The 3 $d^4\theta$ integrals  will
absorb 6 $D$s and 6 $\Dbar$s, leaving 2 $D$s and 2 $\Dbar$s which will
contribute to $\Pi_{1\over2}$ or $\Pi_0$. We are only interested in
knowing  the contribution to $D^2\Dbar^2$, so from this point on we can
treat the $D$s and $\Dbar$s as exactly anticommuting.  We now use
integration by parts  to arrange for each loop to contain 2 $D$s and 2
$\Dbar$s, and integrate the remaining 2 $D$s and 2 $\Dbar$s onto an
external line, writing them in the form $D^2\Dbar^2$. We can now do the
$\theta$ integrals, leaving us with a momentum integral. We evaluate the
momentum  integral, subtract its subdivergences using minimal
subtraction and finally obtain the divergent contribution to
$D^2\Dbar^2$. The reader might like to check  the extent to which this
trick simplifies the one-loop (or even two-loop)  calculation.  The
disadvantage, obviously, is that we lose the nice check afforded  by the
cancellation of the $\Pi_0$ terms. 

The results calculated according to the above procedure 
for each diagram in Fig.~1 are shown in Table~1. The
momentum integrals for each diagram can be expressed  (using integration
by parts) in terms of a basic set depicted in Fig.~2, which can be
evaluated with their subtractions once and for all. In Fig.~2, a dot on a 
line represents a squared propagator, and arrows represent linear momentum 
factors in the numerator, the momenta which 
correspond to a pair of arrows being contracted together.
Each diagram in 
Fig.~2 represents a Feynman integral of logarithmic overall divergence; 
there are no external lines because we have set the external momentum
zero. This means that at least one propagator must be given a mass,
since otherwise  there is no scale defined and  of course dimensional
regularisation cannot be employed: the overall  logarithmic ultra-violet
and infra-red divergences would cancel.  (Alternatively  an external
momentum may be ``threaded'' in an arbitrary way through the  diagram).
Also, any explicit infra-red divergence corresponding to  a squared
propagator must also be regularised by introducing a mass;  such a  mass
will often serve to define the scale, too.  Sometimes an alternative
$\delta$-function infra-red regulator  can be useful\ref\gkz{
M.T.~Grisaru, D.I.~Kazakov and D.~Zanon, \npb 287 (1987) 189\semi
K.G.~Chetyrkin and F.V.~Tkachov, \plb 114 (1982) 133}, whereby  instead
of   
\eqn\irone{{1\over{(k^2)^2}}\to {1\over{(k^2 +
m^2)^2}}\quad\hbox{or}\quad {1\over{k^2(k^2 + m^2)}}} one has
\eqn\irtwo{{1\over{(k^2)^2}}\to {1\over{(k^2)^2}} +
{2\over{4-d}}\delta^{(4)}(k).}
(Note that we have chosen to perform the
diagrammatic calculation in Euclidean space, for which bookkeeping of factors
of $i$ etc. is easier.) 
After subtraction of ultra-violet
subdivergences, the result for each diagram is independent both of the
means by which a scale is introduced and the method used to 
regulate infra-red divergences. (Note that this statement is true only 
of the {\it subtracted\/} diagram.) The second
column of Table~1 shows the expression for each momentum integral  in
terms of those in Fig.~2.  (The presence of a zero in this column may indicate
that the $D$-algebra for the diagram 
gave zero; or alternatively, that the momentum integral
reduced to a ``factorised'' form which 
can be shown\ref\us{I.~Jack, D.R.T.~Jones and N.~Mohammedi, \npb322 (1989) 431}
to produce no simple pole after subtraction.) 
Each diagram in Fig.~1 also corresponds to a
product of group matrices and Yukawa couplings which may be expressed in
terms of basic invariants, and which are given in the third column of 
Table~1 (including also the symmetry factor for the diagram). 
The invariants $X_1$, $X_2$, $X_3$ and $X_4$
are given by
\eqn\inv{\eqalign{X_1&=g^2Y^{klm}P^n{}_lC(R)^p{}_mY_{knp}=g^2\tr\left[S_4
C(R)\right]\cr 
X_2&=g^4Y^{klm}C(R)^n{}_lC(R)^p{}_mY_{knp}=g^4\tr\left[S_1 C(R)\right],\cr
X_3&=g^4\tr[PC(R)^2], \qquad X_4=g^2\tr[P^2C(R)].\cr}}  (See Appendix A for
the definitions of $S_1$, $S_4$ and other notational  details).  Upon
adding all the results, we find 
\eqn\div{<VV>_{\rm pole}=
2r^{-1}\left\{(2A-C)(X_1+2X_3-2g^6Q\tr[C(R)^2])+(2B+C-2D)X_4\right\}.} 
We note that all
contributions involving the diagram $E$ have cancelled; this will be
important, since $E$ is the only diagram out of our basic set in Fig.~2
whose simple pole involves  $\zeta(3)$. We also note that the invariant
$X_2$ does not appear in the  final result, and therefore $<VV>_{\rm
pole}$ (and hence $\beta_g$) vanish in the one-loop finite 
case, when $P=Q=0$.\foot{Of course there is no non--trivial one-loop 
finite abelian theory; we are here really speaking of the corresponding terms 
in the full non-abelian case.} It is interesting  that we can draw this 
conclusion
without evaluating any of the Feynman  integrals.  The $\beta$-function
is derived from the simple poles in the subtracted  diagrams. We find
\eqn\poles{\eqalign{A_{\rm simple}={4\over3}{1\over{\lllf\ep}}, &\qquad
B_{\rm simple}=-{2\over3}{1\over{\lllf\ep}},\cr C_{\rm
simple}={2\over3}{1\over{\lllf\ep}}, &\qquad  D_{\rm
simple}=-{2\over3}{1\over{\lllf\ep}},\cr}} 
and hence we have (note that with our conventions, 
at $L$ loops $\beta_g^{(L)}$ 
differs from the corresponding  simple pole contribution to 
$<VV>_{\rm pole}$ by a factor of $\frak{gL}{4}$)
\eqn\bthree{\lllf\beta_g^{(3)DRED}= r^{-1}g\left\{3X_1+6X_3+X_4-6g^6Q\tr
[C(R)^2]\right\},} where $r=\delta_{AA}$. On the other hand, the NSVZ
result for $\beta_g^{(3)}$ in the abelian case, obtained by setting
$C(G)=0$ in Eq.~\russa, is simply
$\beta_g^{(3)NSVZ}=-2{g^3\over{\lf}}r^{-1}\tr[\ga^{(2)}C(R)]$ , and so
we have \eqn\russb{\lllf\beta_g^{(3)NSVZ}=r^{-1}g\left\{2X_1+4X_3-4g^6Q\tr
[C(R)^2]\right\}.} We note that $X_4$ appears only in the DRED result
Eq.~\bthree, while the other terms appear in the same combination
$X_1+2X_3-2g^6Q\tr[C(R)^2]$ in both Eqs.~\bthree\ and \russb. We also note
that the same combination already appeared in Eq.~\div\ before
evaluating the momentum integrals. 

We now wish to show that the two results Eqs.~\bthree\ and \russb\ are in fact 
related by a coupling constant redefinition, equivalent to a change of 
renormalisation scheme. In general, a redefinition $\delta g$ of $g$ induces a
change $\delta \beta_g$ of $\beta_g$ given (to lowest order in $g$) by
\eqn\redefa{\delta\beta_g (g,Y) = \left[\beta_Y. {{\pa}\over{\pa Y}} 
+ \beta_Y^* . {{\pa}\over{\pa Y^*}} 
  + \beta_g . {{\pa}\over{\pa g}}\right]\delta g
- \delta g . {{\pa}\over{\pa g}} \beta_g.}

In particular, if we choose 
\eqn\redefb{\delta g=-(\lf)^{-2}\frak{1}{2}r^{-1}g^3\tr \left[PC(R)\right],}
then the resulting $\delta\beta_g$ is given by
\eqn\change{\lllf\delta\beta_g=r^{-1}g\left(-X_1-2X_3-X_4+2g^6Q\tr[C(R)^2]
\right).}
We easily see that then
\eqn\russad{\beta^{(3)NSVZ}_g = \beta^{(3)DRED}_g + \delta\beta_g.}
It is obvious from this analysis that, as we already mentioned in the 
Introduction,  it is quite non-trivial that $\beta^{(3)NSVZ}_g $ and 
$\beta^{(3)DRED}_g $
may be related in this way.  Hence this provides a strong check on the validity
of the NSVZ result. 

In these considerations we explored redefinitions of the gauge coupling $g$ 
only; what of redefinitions of the Yukawa couplings $Y^{ijk}$? 
Under an arbitrary redefinition
\eqn\yuka{Y^{ijk} \to (Y^{ijk})'}
we have at once that 
\eqn\yukb{
{\beta_g}' (Y', g) = \beta_g (Y, g) \quad\hbox{and}\quad
\gamma' (Y', g) = \gamma (Y,g).}
It follows that if $\beta_g (Y, g) $ and $\gamma (Y,g)$ satisfy Eq.~\russa, 
then ${\beta_g}' (Y, g)$ and $\gamma' (Y, g)$ do likewise. 
This means that a redefinition $Y\to Y'$ has no effect on the question of 
whether ($\beta_g$, $\gamma$) obey the NSVZ condition. 
This does not mean that redefinitions of $Y$ have no significance, however; 
for example, in Ref.~\ref\jjn{I.~Jack, D.R.T.~Jones and C.G.~North,
\npb 473 (1996) 308}, 
we showed how to construct a redefinition of $Y$ corresponding to a 
change to a scheme 
such that $\gamma^{(3)} = 0$ for a one-loop finite theory. We will return to 
this redefinition in the next section.  
 
We turn now to the non-abelian case. The crucial observation is that 
$\delta g$ as defined in Eq.~\redefb\ {\it does not vanish\/} for 
an $N=2$ theory in general (though it does in the abelian case, 
as may be easily verified). There is, however, an obvious generalisation 
of it to the non-abelian case, to wit 
\eqn\tlfbb{
\delta g = (\lf)^{-2}\frak{1}{2}g^3 
\left[r^{-1}\tr\left[PC(R)\right]-g^2QC(G)\right] 
}
where we have reversed the overall sign (compared to Eq.~\redefb) 
because we plan to use this $\delta g$ to go back from 
$\beta^{(3)NSVZ}_g$ to $ \beta^{(3)DRED}_g$. It is easy to verify that 
Eq.~\tlfbb\ leads to $\delta g = 0$ in the $N=2$ case. Is this the only 
possible extension of $\delta g$ to the non-abelian case? We are constrained 
by the following requirements:
\item{(1)}
$\delta g = 0$ for a one-loop finite theory. This is because we know that both 
$\beta^{(3)NSVZ}_g$ and $ \beta^{(3)DRED}_g$ vanish in the one-loop finite 
case. This is manifest in the NSVZ case, and was explicitly 
verified in Ref.~\pwb\  for the DRED case. (See also Ref.~\gmz.) 
In the three-loop case the relevant $\beta$-functions in 
Eq.~\redefa\ are one-loop. Since these vanish in this case, to produce a
vanishing $\delta \beta_g$, $\delta g$ must vanish also.
\item{(2)}
$\delta g = 0$ for a $N=2$ theory. In the $N=2$ case
we know that, as discussed earlier, $\beta^{(3)NSVZ}_g= \beta^{(3)DRED}_g = 0$. 
Clearly we therefore require $\delta \beta_g=0$; $\delta g=0$ 
ensures this, since the operator $\left[\beta_Y. {{\pa}\over{\pa Y}} 
+ \beta_Y^* . {{\pa}\over{\pa Y^*}} 
  + \beta_g . {{\pa}\over{\pa g}}\right]$ just corresponds to 
multiplication by a factor in the $N=2$ case. 
\item{(3)}
Eq.~\redefb\ must hold in the abelian case.
\item{(4)}
The resulting terms in $\delta\beta_g$ must correspond to possible 
1PI Feynman graphs.

It is easy to convince oneself that Eq.~\tlfbb\ represents the only 
possible transformation (up to an overall constant, which we have 
fixed by the abelian calculation). 
Our result for  $\beta^{(3)DRED}_g$ in the non-abelian case 
is therefore\jjnb:

\eqn\russaf{\eqalign{
\lllf\beta^{(3)DRED}_g &= r^{-1}g\bigl\{3X_1+6X_3+X_4-6g^6Q\tr[C(R)^2]\cr
& -4g^4C(G)\tr[PC(R)]\bigr\}
+g^7QC(G)[4C(G)-Q]\cr
&=
3r^{-1}g^3Y^{ikm}Y_{jkn}P^n{}_mC(R)^j{}_i
+6r^{-1}g^5\tr\left[PC(R)^2\right]\cr&+r^{-1}g^3\tr\left[P^2C(R)\right]
-6r^{-1}Qg^7\tr\left[C(R)^2\right] - 4r^{-1}g^5C(G)\tr\left[PC(R)\right]\cr&
+g^7QC(G)\left[4C(G)-Q\right].\cr}}  
In the general case, no-one has explicitly computed $\beta^{(3)DRED}_g$; 
there does, however exist a calculation in the 
special case of a theory without any chiral superfields by 
Avdeev and Tarasov\ref\avdeev{L.V.~Avdeev and O.V.~Tarasov \plb 112 (1982) 
356}. 
For this special case it is easy to see from Eq.~\russaf\  that we obtain 
\eqn\russag{
\lllf\beta^{(3)DRED}_g = -21g^7C(G)^3}
which precisely agrees with the result of Ref.~\avdeev. 
Note that from Eq.~\russab\ we have that 
\eqn\russah{
\lllf\beta^{(3)NSVZ}_g = -12g^7C(G)^3.}
This difference was first remarked upon in Ref.~\tim. 
The fact that we successfully reproduce Eq.~\russag\ is an excellent check 
of both our abelian calculation and our coupling constant redefinition. 
It is intriguing to note that this redefinition, as defined in 
Eq.~\tlfbb, can be written:
\eqn\russak{
\delta g = -\frak{1}{4}\beta^{(2)}_g.}
This of course suggests the possibility of generalising 
$\delta g$ to all orders, and hence deriving the all-orders $\beta^{(3)DRED}_g$.
 We have been unable to  provide such a construction, however. In the 
next section we proceed to four loops to test whether the simplicity of our 
result Eq.~\russak\ is sustained. 

\newsec{The four-loop calculation}

In the last section we were able to show that the coupling constant 
redefinition $\delta g$ relating $\beta^{(3)DRED}_g$ to $\beta^{(3)NSVZ}$ 
uniquely determined (up to an overall constant) 
without any further calculation. Unfortunately 
this property does not extend to four loops, as is easily seen as follows. 
From Eq.~\russaf\ let us rewrite $\beta^{(3)DRED}_g$ in the form 
\eqn\foura{\beta^{(3)DRED}_g = -4\Delta_1 + 3\Delta_2 + \Delta_3}
where
\eqna\fourb$$\eqalignno{
\lllf\Delta_1&= g^5 C(G)\left[r^{-1}\tr[PC(R)] - g^2QC(G)\right]&\fourb a\cr
\lllf\Delta_2&= r^{-1}\tr\left[g^3S_4C(R) 
-2 g^7 QC(R)^2 + 2 g^5PC(R)^2\right]  &\fourb b\cr
\lllf\Delta_3&= g^3 r^{-1}\tr[P^2C(R)] - g^7 Q^2C(G). &\fourb c\cr}$$
The corresponding formula for $\beta_g^{(3)NSVZ}$ is 
\eqn\fourbb{\beta^{(3)NSVZ}_g = -4\Delta_1 + 2\Delta_2.} 

The purpose of these decompositions is that each of the $\Delta_i$ represents a 
candidate for $\delta g$ satisfying the requirements we formulated in the 
previous section; and, indeed, the $\Delta_i$ are the only such candidates.  
In particular in the $N=2$ case we have 
$\Delta_1 = \Delta_2 = \Delta_3 = 0$. 
 Therefore we may anticipate that 
to relate $\beta^{(4)DRED}_g$ to $\beta^{(4)NSVZ}_g$ we may need to 
make a redefinition of the form $\delta g = \sum \alpha_i \Delta_i$  
where the $\alpha_i$ are as yet undetermined coefficients. We will also,
 of course, need to take into account the effect on the four-loop 
$\beta$-functions of the $O(g^5)$ redefinition discussed in the last section. 
It is immediately clear that we shall be unable to determine the coefficient
$\alpha_1$ from an abelian calculation, since $\Delta_1$ vanishes in the 
abelian case.
Using Eq.~\redefa, we obtain the leading-order change in $\beta_g$ due 
to each of the $\Delta_i$, as follows: 
\eqn\fourc{\eqalign{
\llllf\delta\beta_g (\Delta_1) &=
2g^5 C(G)r^{-1} \Bigl\{ \tr\left[ S_4 C(R)\right] + \tr[P^2C(R)] 
+ 2g^2\tr[PC(R)^2] 
\cr&+ g^2Q\tr[PC(R)]-2g^4Q\tr[C(R)^2]\Bigr\} -4g^9Q^2C(G)^2\cr}}

\eqn\fourd{\eqalign{
\llllf\delta\beta_g (\Delta_2)&= 
2g^3r^{-1}\tr \left[\left(S_7+2S_8 +Y^*S_4Y\right)C(R)+S_5 P\right]\cr& 
+4g^5r^{-1}\tr\left[\left\{S_4C(R)+S_5 
+P^2C(R)\right\}C(R)\right]\cr
&+4Qg^7r^{-1}\tr\left[PC(R)^2 -  S_1 C(R) \right]
+8g^7r^{-1}\tr[PC(R)^3]\cr
&-8Qg^9r^{-1}\tr\left[ C(R)^3 +Q C(R)^2\right]\cr}}

\eqn\foure{\eqalign{
\llllf\delta\beta_g (\Delta_3) &=4g^3r^{-1}\tr\left[P^3C(R) +2g^2P^2C(R)^2
-2g^4QPC(R)^2 +S_5 P\right]\cr&-4g^9Q^3C(G).\cr}}

It is straightforward to verify that each of the 
$\delta\beta_g (\Delta_i)$ vanishes for $N=2$ supersymmetry; manifestly they 
also vanish in a one-loop finite theory. We must be careful about the logic 
of our procedure here, because we know that $\gamma^{(3)DRED}$ does not vanish 
in a one-loop finite theory
\ref\parkes{A.J.~Parkes, \plb 156 (1985) 73}, 
so there is no reason to expect that 
$\beta_g^{(4)DRED}$ will either (and, indeed, it does not). There is, however, 
a redefinition $Y\to Y + \delta Y$ corresponding to a change to a 
renormalisation scheme (DRED$'$, say) such that  
$\gamma^{(3)DRED'}$ does vanish in the one-loop finite case\jjn. 
In this scheme $\beta_g^{(4)DRED'}$ will also vanish, 
relying on the theorem\pwb\gmz\ 
that in an $n$-loop finite theory $\beta_g^{(n+1)} = 0$. Suppose that we have 
found a $\delta g$ which transforms $\beta_g^{(4)DRED}$ to the NSVZ form
$\beta_g^{(4)NSVZ}$ (i.e. consistent with Eq.~\russa). (We must take into 
account here the fact that $\gamma^{(3)NSVZ}$ differs from $\gamma^{(3)DRED}$,
owing to the redefinition of $g$ which takes us from DRED to the NSVZ form at 
the three-loop level--see later.)
Applying the above $\delta Y$ to $\gamma^{(3)NSVZ}$ and $\beta_g^{(4)NSVZ}$
yields a new $\gamma^{(3)NSVZ'}$ and $\beta_g^{(4)NSVZ'}$ which are also
of the NSVZ form, recalling that a redefinition of $Y$ 
does not affect whether or not $\beta_g$ satisfies Eq.~\russa. Our redefinition 
$\delta g$ also transforms $\beta_g^{(4)DRED'}$ into $\beta_g^{(4)NSVZ'}$,
which both vanish in the one-loop finite case, and hence we conclude that
$\delta g$ must itself vanish in the one-loop finite case.  

From Eq.~\russa, we have that 
\eqn\fourf{\eqalign{\llllf\beta^{(4)NSVZ}_g &=
8g^9 Q C(G)^3 -8g^7 C(G)^2 r^{-1} \lf \tr\left[ \ga^{(1)}C (R)\right]
\cr&-4g^5 C(G) r^{-1} \llf \tr\left[ \ga^{(2)}C (R)\right]\cr
&-2g^3  r^{-1} \lllf \tr\left[ \ga^{(3)NSVZ}C (R)\right].\cr}}
Now in this equation, $\gamma^{(1)}$ and $\gamma^{(2)}$ are unambiguous, 
being defined in Eq.~\Ed\ and \Au{b}\ respectively, but we must be careful 
about $\gamma^{(3)}$. 

From Ref.~\jjn\ we have the result for $\gamma^{(3)DRED}$:
\eqn\tlfa{\eqalign{\lllf\ga^{(3)DRED} = 
& \kappa \bigl\{g^6 \left[ 12 C(R) C(G)^2 - 2 C(R)^2 C(G) -10 C(R)^3   
-4 C(R)\Delta (R)\right]\cr   &
+ g^4\left[ 4C(R) S_1 - C(G)S_1 + S_2 -5 S_3\right] 
+ g^2Y^* S_1 Y +  \frak{1}{4}M\cr 
&+  g^2\left[C(R)S_4 -2S_5 -S_6 \right] - g^4
\left[PC(R)C(G) +5PC(R)^2\right]\cr& 
+4g^6QC(G)C(R) \bigr\}
 +2Y^*S_4 Y - \half S_7 - S_8 +g^2\left[ 4C(R)S_4  + 4S_5\right]\cr
& + 
g^4\left[8C(R)^2 P  -2Q C(R) P - 4QS_1 
- 10r^{-1}{\rm Tr}\left[PC(R)\right]C(R)\right]\cr 
&  +g^6\left[2Q^2C(R)-8C(R)^2 Q  + 10QC(R)C(G)\right]
 \cr}}
where $\kappa = 6\zeta (3)$. Group theoretic factors are defined in 
Appendix A.

Now in the previous section we constructed a $\delta g$ that related 
$\beta^{(3)DRED}_g$ to $\beta^{(3)NSVZ}_g$, Eq.~\tlfbb. As we mentioned earlier,
this redefinition affects $\gamma$ too, so that 
\eqn\delgam{\eqalign{
\gamma^{(3)NSVZ} &= \gamma^{(3)DRED} - 4g\delta g C(R)(\lf)^{-1}\cr
&= \gamma^{(3)DRED}-2 (\lf)^{-3}g^4 
\left[r^{-1}\tr\left[PC(R)\right]-g^2QC(G)\right]C(R).\cr 
}}
This completes the definition of $\beta^{(4)NSVZ}_g$.  We anticipate 
that it will be related to  $\beta^{(4)DRED}_g$ as follows:

\eqn\betaf{\beta^{(4)DRED}_g = \beta^{(4)NSVZ}_g + 
\sum_i \alpha_i \delta\beta_g (\Delta_i ) + \Omega}
where $\Omega$ is the change in $\beta^{(4)}_g$ due to the redefinition 
in Eq.~\tlfbb, 
\eqn\delomega{ \Omega = \left[\beta^{(2)}_Y. {{\pa}\over{\pa Y}} 
+ \beta^{*(2)}_{Y} . {{\pa}\over{\pa Y^*}} 
  + \beta^{(2)}_g . {{\pa}\over{\pa g}}\right]\delta g
- \delta g . {{\pa}\over{\pa g}} \beta^{(2)}_g.}
(There are, of course, terms of $O\left((\delta g)^2\right)$, 
but these are $O(g^{11}$) and 
hence affect $\beta_g^{(5)}$.)  
Substituting for $\beta^{(2)}_Y$ and $\beta^{(2)}_g$ we obtain:
\eqn\delomegab{\eqalign{\llllf\Omega &= g^3r^{-1}\bigl\{
2Qg^4\tr[PC(R)^2 + 2g^2C(R)^3] - 2g^2\tr[P^2C(R)^2 + 2g^2PC(R)^3]\cr
&+ 2g^4Q\tr\left[ S_1 C(R)\right] 
- \tr[Y^*S_4YC(R)]\cr
&-\tr\left[2g^2\left( S_5 + S_4C(R)\right)C(R)
 + S_5 P\right]\bigr\}.\cr}}
Given an explicit calculation of $\beta^{(4)DRED}_g$, we would 
be able to test the validity of the construction Eq.~\betaf; and 
if it proved to be valid, determine the $\alpha_i$. This 
calculation is beyond our strength; we have, however, performed a 
partial calculation consisting of the contributions to $\beta^{(4)DRED}_g$   
in the abelian theory of $O(g^3Y^6)$. 
From Eqs.~\fourc--\foure\ it is clear that this will enable 
us both to test whether our construction works with regard to such terms and
 also to determine $\alpha_2$ and $\alpha_3$, thus fixing $\beta^{(4)DRED}_g$ 
apart
from a single unknown parameter ($\alpha_1$). The calculations were 
performed using the methods explained earlier in the three-loop case. The
relevant diagrams are shown in Fig. 3, and the results are given in Table~2.
As before, the momentum integrals can be expressed in terms of a convenient
basis, depicted in Fig.~4. The simple poles for each of these (subtracted)
diagrams are given by
\eqn\momfour{\eqalign{F_{\rm simple}
&={1\over{\llllf\epsilon}}\left({1\over2}-\zeta(3)\right),\phantom{-}
\quad G_{\rm simple}=2{1\over{\llllf\epsilon}}\biggl(-1+\zeta(3)\biggr),\cr
H_{\rm simple}&={1\over{\llllf\epsilon}}\left(-{5\over6}+\zeta(3)\right),
\quad I_{\rm simple}={5\over2}{1\over{\llllf\epsilon}},\cr
\quad J_{\rm simple}&=-{2\over3}{1\over{\llllf\epsilon}},
\quad\hskip 1.6cm  K_{\rm simple}=-{5\over6}{1\over{\llllf\epsilon}},\cr
\quad\hskip 1.6cm  L_{\rm simple}&=-{1\over2}{1\over{\llllf\epsilon}},
\quad\hskip 1.6cm M_{\rm simple}={11\over6}{1\over{\llllf\epsilon}},\cr
\quad\hskip 1.6cm N_{\rm simple}&=-{1\over6}{1\over{\llllf\epsilon}},
\quad\hskip 1.6cm P_{\rm simple}=-{1\over6}{1\over{\llllf\epsilon}},\cr
\quad\hskip 1.6cm Q_{\rm simple}&=-{2\over3}{1\over{\llllf\epsilon}},
\quad\hskip 1.6cm R_{\rm simple}=-{1\over2}{1\over{\llllf\epsilon}}.\cr}}
It is interesting to note that $\zeta(3)$ dependence only appears in those
diagrams which contain a ``figure-of-eight'' sub-diagram. This could 
perhaps be explained by invoking recent proposals which relate  
the appearance of transcendental numbers in Feynman integrals to knot 
theory\ref\knot{D.~Kreimer, \plb354 (1995) 117\semi
D.~Kreimer, hep-th/9412045\semi
D.~Broadhurst and D.~Kreimer, hep-ph/9504352\semi
D.~Kreimer, hep-ph/9505236\semi
D.~Broadhurst, J.A.~Gracey and D.~Kreimer, hep-th/9607174}. On adding the
results for all the diagrams, we find
\eqn\betaDRED{\eqalign{
\llllf\beta_g^{(4)DRED}&=
\frak{1}{3}g^3r^{-1}\tr\left[\left(2P^3-2S_8-19Y^*S_4Y-S_7\right)C(R)\right]\cr
&-\frak{1}{2}\kappa g^3r^{-1}\tr[MC(R)]
-\frak{5}{3}g^3r^{-1}\tr [S_5 P]\cr
&+\cdots \quad(\hbox{terms involving $C(G)$, 
and of order $g^5Y^4$, $g^7Y^2$ and $g^9$}.)\cr}}
The coefficient of the $\tr[MC(R)]$ term was derived not by direct calculation, 
but by exploiting the fact that, as we observed earlier, 
the coupling constant redefinition which 
makes $\gamma^{(3)}$ vanish in the one-loop finite case should also make
$\beta_g^{(4)}$ vanish. It was shown in Ref.~\jjn\ that the redefinition
\eqn\Yredef{\llf\delta Y^{ijk}=
\frak{1}{4}\kappa Y^{ilm}Y^{jpq}Y^{krs}Y_{lpr}Y_{mqs}
+O(g^2)}
makes $\gamma^{(3)}$ vanish in the one-loop finite case. Since a change 
$\delta Y$ induces a leading-order change in $\beta_g$ of the form
\eqn\dbetag{\eqalign{
\delta \beta_g&=
-\left[\delta Y.{\partial\over{\partial Y}}
+\delta Y^*.{\partial\over{\partial Y^*}}\right]\beta_g^{(2)}\cr 
&=(\lf)^{-4}r^{-1}g^3C(R)^i{}_j(Y^{jkl}\delta Y_{ikl}+\delta Y^{jkl} Y_{ikl}),}}
we can deduce the coefficient of $\tr[MC(R)]$ to be as given. This indirect
deduction was necessitated because the direct calculation of this 
coefficient proved very involved. One might feel slightly uneasy because
it is not clear that the theorem that $\beta_g^{(n+1)}$ vanishes in an $n$-loop
finite theory should be valid in an arbitrary renormalisation scheme. To
allay these doubts, we have explicitly computed the coefficients of 
$\tr[C(R)^2\Delta(R)]$ and $\tr[Y^*S_1YC(R)]$ in $\beta_g^{(4)DRED}$ in the 
one-loop finite case, and checked that they agree with those obtained by the 
same indirect argument.  

Now comparing the result Eq.~\betaDRED\ with Eqs.~\fourc--\fourf\ and 
\delomegab,
we see that we require
\eqn\alphs{
\alpha_2=-\frak{2}{3},\qquad \alpha_3=\frak{1}{6}.}
We note that it is non-trivial that a solution exists at all for $\alpha_2$ and
$\alpha_3$, since we need to reproduce six terms in Eq.~\betaDRED\ and 
we have only two free
parameters $\alpha_2$ and $\alpha_3$ to adjust. Note, however, that the 
coefficient of $\tr[MC(R)]$ automatically satisfies Eq.~\fourf, irrespective of 
the values of $\alpha_2$ and $\alpha_3$; this is 
essentially guaranteed, as we know that we can redefine $Y$ so that the $M$ 
terms vanish in both $\gamma^{(3)}$ and $\beta_g^{(4)}$ (hence trivially
satisfying Eq.~\fourf\ as far as these terms are concerned), and we also know,
as mentioned earlier, that redefinitions of $Y$ have no effect on whether
Eq.~\fourf\ is satisfied. 

As we have indicated, several miracles were required to facilitate our 
construction. It is perhaps disappointing, however, that it remains 
unclear how the redefinition we have found generalises to higher orders. From 
Eq.~\russak, it was tempting to conjecture that the transformation 
$\delta g = \sum \alpha_i \Delta_i$ might have been proportional to 
$\beta_g^{(3)DRED}$ or $\beta_g^{(3)NSVZ}$ 
but it is easy to see from Eqs.~\foura\ and \fourbb\
that this doesn't work. 

The final complete result is
\eqn\final{\eqalign{
\llllf\beta_g^{(4)DRED}&=g^3r^{-1}\tr\left[\left\{-\frak{1}{2}\kappa M
-\frak{19}{3}Y^*S_4Y+\frak{2}{3}P^3-\frak{1}{3}S_7-\frak{2}{3}S_8\right\}C(R)
-\frak{5}{3}S_5 P\right]\cr
&+g^5r^{-1}\tr\Bigl[\Bigl\{-2\kappa Y^*S_1Y-(2\kappa+\frak{38}{3})C(R)S_4
+(4\kappa-\frak{38}{3})S_5+2\kappa S_6\cr
&-\frak{10}{3}P^2C(R)+(2\alpha_1+4)C(G)S_4+2\alpha_1C(G)P^2\Bigr\}C(R)\Bigr]\cr
&+g^7r^{-1}\tr\Bigl[\Bigl\{-8\kappa C(R)S_1+2\kappa C(G)S_1
-2\kappa S_2+10\kappa S_3\cr&+24r^{-1}\tr[PC(R)]C(R)
+(10\kappa- \frak{76}{3})PC(R)^2+2QPC(R)+\frak{38}{3}QS_1\cr&
+(2\kappa+4\alpha_1+8)C(G)PC(R)+2\alpha_1QC(G)P-8C(G)^2P\Bigr\}C(R)\Bigr]\cr
&+g^9r^{-1}\tr\Bigl[\frak{76}{3}QC(R)^3+\frak{4}{3}Q^2C(R)^2
-(8\kappa+4\alpha_1+32)QC(G)C(R)^2\cr
&-24\kappa C(G)^2C(R)^2+4\kappa C(G)C(R)^3+20\kappa C(R)^4
+8\kappa\Delta(R)C(R)^2\Bigr]\cr
&+g^9\left[8QC(G)^3-4\alpha_1Q^2C(G)^2 -\frak{2}{3}Q^3C(G)\right]\cr}}

We see that in a one-loop finite theory ($P=Q=0$), 
we have $\beta_g^{(4)DRED}\neq 0$; but, as noted earlier,
it can be transformed to zero 
by means of the redefinition $Y\to Y + \delta Y$, where $\delta Y$ is 
the transformation that makes $\gamma^{(3)DRED}$ vanish in such a theory. 
We gave the $O(Y^5)$ term in this $\delta Y$ in Eq.~\Yredef; 
the full expression is\jjn 
\eqn\tlfc{\eqalign{
(16\pi^2)^3\delta Y^{ijk} &=  
\kappa \bigl\{ \frak{1}{4} Y^{ilm}Y^{jpq}Y^{krs}Y_{lpr}Y_{mqs}
+\half g^2S_1{}^{(i}{}_m Y^{jk)m}\cr   
&+ g^4[C(R)^{(i}{}_m C(R)^j{}_n Y^{k)mn}
-\frak{5}{2} Y^{n(jk}C(R)^{i)}{}_m C(R)^m{}_n - \Delta (R) Y^{ijk}\cr
&-\half C(G) C(R)^{(i}{}_m Y^{jk)m} + 3 C(G)^2 Y^{ijk}]\bigr\}.\cr
}}
This redefinition defines the DRED$'$ scheme which we introduced previously; 
correspondingly, the same redefinition applied to 
$\beta_g^{(4)NSVZ}$, as given in Eq.~\fourf, will define the NSVZ$'$ scheme 
for which (given $P=Q=0$) $\beta_g^{(4)NSVZ'} = 0$ likewise
(as can readily be checked using Eq.~\dbetag).  

\newsec{Conclusions}

We have explicitly constructed the coupling constant redefinitions that 
relate $\beta_g^{NSVZ}$ to $\beta_g^{DRED}$ up to and including four
loops,  except for one undetermined parameter. The fact that this
construction was  possible  demonstrates that the renormalisation scheme
in which the  $NSVZ$ form is valid is perturbatively related to the
conventional  DRED scheme. As a by-product of our investigations we have
obtained  $\beta_g^{(3)DRED}$ and $\beta_g^{(4)DRED}$ for a general 
non-abelian theory, 
except for the dependence of $\beta_g^{(4)DRED}$ on the same undetermined
parameter. 

The fixed points of $\beta_g^{NSVZ}$, which satisfy the equation 
\eqn\conca{2r^{-1}\tr\left[\ga C(R)\right] = Q} are significant in dual
gauge theories. Now the existence of a fixed point is  preserved under a
coupling constant redefinition $g\to g'(g)$, as long as the  function
$g'(g)$ is differentiable at the fixed point. Our demonstration that 
the NSVZ and DRED schemes are perturbatively related  therefore suggests
that there is a fixed point of $\beta_g^{DRED}$ corresponding to any
fixed point of $\beta_g^{NSVZ}$. It would have been interesting had we 
been able to construct the $DRED\leftrightarrow NSVZ$ redefinition to 
all orders;  unfortunately,  however, to the extent that we have pursued
the perturbative form, there  is no indication of an all-orders result.
It is tantalising, in this regard, that in the development of 
background-field covariant superfield Feynman rules
\ref\grisz{M.T.~Grisaru and D.~Zanon, \npb 252 (1985) 578}\  it appears
that beyond one loop, loops of $\epsilon$-scalars play a crucial role in
the relevant counter-terms.  Such a loop provides a factor of
$\epsilon$, so that the simple pole in $\epsilon$  thus depends on what
would have been the double pole had the  $\epsilon$-scalars been
ordinary. Given that all higher order poles in $\epsilon$ are determined
 in terms of the simple pole by the 't Hooft consistency conditions
\ref\hooft{G.~'t Hooft, \npb 62 (1973) 444}, one might  have hoped to
proceed to an all-orders construction of $\beta_g^{DRED}$. In this endeavour 
we have not been successful.  

Nevertheless, we feel that the exercise has been worthwhile. We have
demonstrated  beyond all reasonable doubt that there does exist a scheme
in which  the $NSVZ$ $\beta$-function is valid.  Our result for
$\beta_g^{(3)DRED}$, in conjunction with the  result  for
$\gamma^{(3)DRED}$ from Ref.~\jjn, is in any event of  interest
phenomenologically \ref\fjj{P.M.~Ferreira, I.~Jack and  D.R.T.~Jones,
hep-ph/9605440},  especially, perhaps, in post-post-modern theories with
additional matter content (see for example \ref\dine{M.~Dine, Y.~Nir and 
Y.~Shirman, hep-ph/9607397}).

\bigskip\centerline{{\bf Acknowledgements}}\nobreak

Part of this work was done during a visit by two of us (IJ and TJ) to
the Aspen Center for Physics. TJ also thanks the physicists at 
the University of Colorado, Boulder for hospitality. Financial 
support was provided by PPARC, the Royal Society and the University of 
Colorado.  
IJ and CGN were supported by PPARC via an Advanced Fellowship and a
Graduate Studentship respectively.

\appendix{A}{Notation and Conventions}

Here we give some details concerning our notation and various 
useful formulae. With our conventions, the $D$-algebra in Minkowski space 
is 
\eqn\adalg{\eqalign{
\{D_{\alpha},\Dbar_{\dot\alpha}\}&=\frak{1}{2}i\sigma^{\mu}_{\alpha\dot\alpha}
\partial_{\mu},\cr
\{D_{\alpha},D_{\beta}\}&=\{\Dbar_{\alpha},\Dbar_{\beta}\}=0,\cr}}
where $\sigma^{\mu}=(I,\sigma^i)$, $\sigma^i$ being the Pauli matrices.
It is then easy to verify that the projection operators
\eqn\pia{\eqalign{
\Pi_0&=D^2\Dbar^2+\Dbar^2D^2,\cr
\Pi_{1\over2}&=-2D^{\al}\Dbar^2D_{\al}\cr}}
satisfy
\eqn\pib{\Pi_{1\over2}+\Pi_0=-\partial^2}
as stated in Section~2. 
We use standard superfield Feynman rules, as described, for example, 
in Ref.~\ref\grsb{M.~Grisaru,
W.~Siegel and M.~Rocek, \npb 159 (1979) 429}, with minor changes
of normalisation consequent upon our conventions in Eq.~\adalg; the only 
unusual feature being the  way we use transversality of the 
vector self-energy to simplify the $D$-algebra, as described in Section 2.  
A useful identity in manipulation of supergraphs 
(which follows immediately from Eqs.~\pia, \pib) is
\eqn\pic{D^2\Dbar^2D^2=-\partial^2D^2, \quad \Dbar^2D^2\Dbar^2
=-\partial^2\Dbar^2.}
Gauge invariance of the superpotential means that the Yukawa couplings 
$Y^{ijk}$ satisfy the equation
\eqn\aone{Y^{m(ij} R_A^{k)}{}_m = 0}
From this identity it is straightforward to prove that:
\eqn\atwo{Y^{imn} R_A^{j}{}_m R_A^{k}{}_n
=\half\left[ Y^{mjk}C(R)^{i}{}_m - Y^{imk}C(R)^{j}{}_m
- Y^{ijm}C(R)^{k}{}_m\right]. 
}
Eq.~\aone\ is rather similar to momentum conservation at a three-point 
vertex; correspondingly, Eq.~\atwo\ is analogous to the simple 
identity 
\eqn\athree{ p.q = \half\left[(p+q)^2 - p^2 - q^2\right]}
Eq.~\atwo\ is very useful in dealing with the group theoretic 
factors. It is always possible, in the calculations we have performed,
to push $R_A$'s around so as to produce the quadratic 
Casimir $C(R)$ (though there is no reason to expect this property to
persist to arbitrary orders). In this process, much labour involving 
dummy indices is avoided by the adoption 
of a diagrammatic notation where $Y$ or $Y^*$ are represented by 
vertices, index contractions $\delta^i{}_j$ by   
propagators, and $R_A$ by ``mass insertions''.

We use the following definitions:
\eqna\tlfb$$\eqalignno{
S_1 ^i{}_j &= Y^{imn}C(R)^p{}_m Y_{jpn}&\tlfb a\cr
(Y^*S_1 Y)^i{}_j &= Y^{imn}S_1{}^p{}_m Y_{jpn}&\tlfb b\cr
S_2 ^i{}_j &= Y^{imn}C(R)^p{}_m C(R)^q{}_n Y_{jpq}&\tlfb c\cr
S_3 ^i{}_j &= Y^{imn}(C(R)^2)^p{}_m Y_{jpn}&\tlfb d\cr
S_4 ^i{}_j &= Y^{imn}P^p{}_m Y_{jpn}&\tlfb e\cr
(Y^*S_4 Y)^i{}_j &= Y^{imn}S_4{}^p{}_m Y_{jpn}.&\tlfb f\cr
S_5^i{}_j &= Y^{imn}C(R)^p{}_m P^q{}_p Y_{jnq}&\tlfb g\cr
S_6 ^i{}_j &= Y^{imn}C(R)^p{}_m P^q{}_n Y_{jpq}&\tlfb h\cr
S_7 ^i{}_j &= Y^{imn}P^p{}_m P^q{}_n Y_{jpq}&\tlfb i\cr
S_8 ^i{}_j &= Y^{imn}(P^2)^p{}_m Y_{jpn}&\tlfb j\cr
\Delta (R) &= \sum_{\alpha} C(R_{\alpha})T(R_{\alpha})&\tlfb k\cr
M^i{}_j &= Y^{ikl}Y_{kmn}Y_{lrs}Y^{pmr}Y^{qns}Y_{jpq}&\tlfb l\cr}
$$
  In Eq.~\tlfb{k}\ the sum over $\alpha$ is a sum 
over irreducible  representations.

\listrefs

\vfill\eject

\nopagenumbers
$$\vbox{\offinterlineskip
\def\vr{\vrule height 11pt depth 5pt}
\def\vrq{\vr\quad}
\settabs
\+
\vrq Diagram  \quad & \vrq \qquad Integrals
 \quad\qquad\
& \vrq \quad $-8X_2+24X_3+48g^6\tr[C(R)^3]$\quad&
\vr\cr\hrule
\+
\vrq Diagram \quad & \vrq \quad
 Integrals \quad\quad
& \vrq \quad Group factor \quad\quad&
\vr\cr\hrule
\+
\vrq a \quad & \vrq  \quad $0$
& \vrq \quad& 
\vr\cr\hrule
\+
\vrq b \quad & \vrq  \quad $C$
& \vrq $2X_4+4X_3-2X_1$\quad&
\vr\cr\hrule
\+
\vrq c  \quad & \vrq  \quad $A$
& \vrq $4X_1$\quad&
\vr\cr\hrule
\+
\vrq d \quad & \vrq  \quad
$B$
& \vrq $2X_4$\quad&
\vr\cr\hrule
\+
\vrq e  \quad &
\vrq  \quad $B$
& \vrq $4X_4$\quad&
\vr\cr\hrule
\+
\vrq f  \quad & \vrq  \quad $2A+2B$
& \vrq $-16X_3$\quad&
\vr\cr\hrule
\+
\vrq g  \quad & \vrq  \quad $A$
& \vrq $-8X_3$\quad&
\vr\cr\hrule
\+
\vrq h \quad & \vrq  \quad $A$
& \vrq $16X_3$\quad&
\vr\cr\hrule
\+
\vrq i  \quad & \vrq  \quad $B$
& \vrq $16X_3$\quad&
\vr\cr\hrule
\+
\vrq j \quad & \vrq  \quad $B$
& \vrq $16X_3$\quad&
\vr\cr\hrule
\+
\vrq k \quad & \vrq  \quad $-2A+{1\over2}C$
& \vrq $8g^6Q\tr[C(R)^2]$\quad&
\vr\cr\hrule
\+
\vrq l \quad &
\vrq  \quad $-{1\over2}A+{1\over2}B$
& \vrq $16g^6Q\tr[C(R)^2]$\quad&
\vr\cr
\hrule
\+
\vrq m \quad & \vrq  \quad $-{1\over2}A+{1\over2}B$
& \vrq $-32g^6Q\tr[C(R)^2]$\quad&
\vr\cr\hrule
\+
\vrq n \quad & \vrq  \quad $B$
& \vrq $8g^6Q\tr[C(R)^2]$\quad&
\vr\cr\hrule
\+
\vrq o \quad & \vrq  \quad $-B-2D$
& \vrq $8X_3+2X_4+8g^6\tr[C(R)^3]$\quad& 
\vr\cr\hrule
\+
\vrq p\quad & \vrq  \quad $0$
& \vrq \quad& 
\vr\cr\hrule
\+
\vrq q  \quad & \vrq  \quad $A$
& \vrq $-8X_2+24X_3+48g^6\tr[C(R)^3]$\quad&
\vr\cr\hrule
\+
\vrq r  \quad & \vrq  \quad $2A+2D$
& \vrq $4X_2-12X_3-24g^6\tr[C(R)^3]$\quad&
\vr\cr\hrule
\+
\vrq s  \quad & \vrq  \quad
$2A-2D-E$
& \vrq $4X_2-12X_3-24g^6\tr[C(R)^3]$\quad&
\vr\cr\hrule
\+
\vrq t  \quad &
\vrq  \quad $A$
& \vrq $8X_3-8X_2+16g^6\tr[C(R)^3]$\quad&
\vr\cr\hrule
\+
\vrq u  \quad & \vrq  \quad $0$
& \vrq \quad&
\vr\cr\hrule
\+
\vrq v  \quad & \vrq  \quad $E$
& \vrq $4X_2-4X_3-8g^6\tr[C(R)^3]$\quad&
\vr\cr\hrule
\+
\vrq w  \quad & \vrq  \quad $2A+B+2D$
& \vrq $8X_3+16g^6\tr[C(R)^3]$\quad&
\vr\cr\hrule
\+
\vrq x  \quad & \vrq  \quad $0$
& \vrq \quad&
\vr\cr\hrule
\+
\vrq y \quad & \vrq  \quad $4A-C-E$
& \vrq $8X_3+16g^6\tr[C(R)^3]$\quad&
\vr\cr\hrule
\+
\vrq z  \quad & \vrq  \quad $A$
& \vrq $-16X_3-32g^6\tr[C(R)^3]$\quad&
\vr\cr\hrule
}$$
\in
{\it \noindent
Table 1: Three-loop contributions to $<VV>_{\rm pole}$ for the 
diagrams in Figure 1. Each contribution is obtained by multiplying the 
simple pole from the momentum integral in the first column by the group
theory factor in the second column, and by $r^{-1}$.}
\out
\vfill\eject
$$\vbox{\offinterlineskip
\def\vr{\vrule height 11pt depth 5pt}
\def\vrq{\vr\quad}
\settabs
\+
\vrq Diagram  \quad & \vrq \qquad Integrals
 \quad\qquad\
& \vrq \quad$16X_3+32g^6\tr[C(R)^3]$\quad&
\vr\cr\hrule
\+
\vrq Diagram \quad & \vrq \quad
 Integrals \quad\quad
& \vrq Group factor&
\vr\cr\hrule
\+
\vrq aa  \quad &
\vrq  \quad $A$
& \vrq $16X_3+32g^6\tr[C(R)^3]$\quad&
\vr\cr
\hrule
\+
\vrq bb \quad & \vrq  \quad $2A+2B$
& \vrq $-32g^6\tr[C(R)^3]$\quad&
\vr\cr\hrule
\+
\vrq cc \quad & \vrq  \quad $B$
& \vrq $16g^6\tr[C(R)^3]$\quad&
\vr\cr\hrule
\+
\vrq dd  \quad & \vrq  \quad $B$
& \vrq $16g^6\tr[C(R)^3]$ \quad&
\vr\cr\hrule
\+
\vrq ee \quad & \vrq  \quad $A$
& \vrq $32g^6\tr[C(R)^3]$\quad&
\vr\cr\hrule
\+
\vrq ff \quad & \vrq  \quad $B+E$
& \vrq $-32g^6\tr[C(R)^3]$\quad&
\vr\cr\hrule
\+
\vrq gg \quad & \vrq  \quad $E$
& \vrq $16g^6\tr[C(R)^3]$\quad&
\vr\cr\hrule
\+
\vrq hh \quad & \vrq  \quad $4A+3B-2D$  
& \vrq $8g^6\tr[C(R)^3]$\quad& 
\vr\cr\hrule
\+
\vrq ii \quad & \vrq  \quad $2B-8D+2E$
& \vrq $8g^6\tr[C(R)^3]$\quad& 
\vr\cr\hrule
\+
\vrq jj \quad & \vrq  \quad $4B+C+2D+E$ 
& \vrq $16g^6\tr[C(R)^3]$\quad&
\vr\cr\hrule
\+
\vrq kk \quad & \vrq  \quad $B+E$
& \vrq $16g^6\tr[C(R)^3]$\quad&
\vr\cr\hrule
\+
\vrq ll \quad & \vrq  \quad $B+E$      
& \vrq $-32g^6\tr[C(R)^3]$\quad& 
\vr\cr\hrule
\+
\vrq mm \quad & \vrq  \quad $B$
& \vrq $32g^6\tr[C(R)^3]$\quad&
\vr\cr\hrule
\+
\vrq nn \quad & \vrq  \quad $A$  
& \vrq $32g^6\tr[C(R)^3]$\quad& 
\vr\cr\hrule
\+ 
\vrq pp \quad & \vrq  \quad $2A+2B$
& \vrq $-32g^6\tr[C(R)^3]$\quad&
\vr\cr\hrule
\+
\vrq qq \quad & \vrq  \quad $B$
& \vrq $-16g^6\tr[C(R)^3]$\quad&
\vr\cr\hrule
\+
\vrq rr \quad & \vrq  \quad $2A-2D$    
& \vrq $-16g^6\tr[C(R)^3]$\quad&
\vr\cr\hrule
\+ 
\vrq ss \quad & \vrq  \quad $A$
& \vrq $-32g^6\tr[C(R)^3]$\quad&
\vr\cr\hrule
\+ 
\vrq tt \quad & \vrq  \quad $A$    
& \vrq $32g^6\tr[C(R)^3]$\quad&
\vr\cr\hrule
\+
\vrq uu \quad & \vrq  \quad $B$    
& \vrq $16g^6\tr[C(R)^3]$\quad& 
\vr\cr\hrule
\+
\vrq vv \quad & \vrq  \quad $0$
& \vrq \quad&
\vr\cr\hrule
\+
\vrq ww \quad & \vrq  \quad $0$
& \vrq \quad&
\vr\cr\hrule
\+
\vrq xx \quad & \vrq  \quad $0$
& \vrq \quad&
\vr\cr\hrule
}$$
\in
{\it \noindent
Table 1 (continued)}
\out
\vfill
\eject
$$\vbox{\offinterlineskip
\def\vr{\vrule height 11pt depth 5pt}
\def\vrq{\vr\quad}
\settabs
\+
\vrq Diagram  \quad & \vrq \qquad Integrals
 \quad\qquad\
& \vrq \quad $2\tr[-Y^*S_4YC(R)+S_5P]$\quad&
\vr\cr\hrule   
\+ 
\vrq Diagram  \quad & \vrq \qquad Integrals
 \quad\qquad\
& \vrq Group Factor&
\vr\cr\hrule
\+
\vrq a \quad & \vrq \qquad $4P-2R+2J$
 \quad\qquad\
& \vrq $\tr[S_7C(R)-2S_5P]$&
\vr\cr\hrule
\+
\vrq b \quad & \vrq \qquad $F+2N$     
 \quad\qquad\
& \vrq $4\tr[P^3C(R)]$&
\vr\cr\hrule
\+
\vrq c \quad & \vrq \qquad $J+2Q-L$    
 \quad\qquad\
& \vrq $2\tr[S_7C(R)]$&               
\vr\cr\hrule
\+
\vrq d \quad & \vrq \qquad $J+2P$    
 \quad\qquad\
& \vrq $8\tr[S_5P]-4\tr[S_7C(R)]$&               
\vr\cr\hrule
\+
\vrq e \quad & \vrq \qquad $-I$ 
 \quad\qquad\
& \vrq $4\tr[Y^*S_4YC(R)]$&             
\vr\cr\hrule
\+
\vrq f \quad & \vrq \qquad $-M$
 \quad\qquad\
& \vrq $2\tr[-Y^*S_4YC(R)+S_5P]$&
\vr\cr\hrule
\+
\vrq g \quad & \vrq \qquad $0$     
 \quad\qquad\
& \vrq $$&            
\vr\cr\hrule
\+ 
\vrq h \quad & \vrq \qquad $0$     
 \quad\qquad\
& \vrq $$&            
\vr\cr\hrule
\+ 
\vrq i \quad & \vrq \qquad $0$     
 \quad\qquad\
& \vrq $$&            
\vr\cr\hrule
\+ 
\vrq j \quad & \vrq \qquad $0$     
 \quad\qquad\
& \vrq $$&            
\vr\cr\hrule
\+ 
\vrq k \quad & \vrq \qquad $-L$     
 \quad\qquad\
& \vrq $-4\tr[S_7C(R)]$&            
\vr\cr\hrule
\+ 
\vrq l\quad & \vrq \qquad $-G$     
 \quad\qquad\
& \vrq $2\tr[(P^3-S_8)C(R)]$&            
\vr\cr\hrule
\+ 
\vrq m \quad & \vrq \qquad $-K$     
 \quad\qquad\
& \vrq $2\tr[S_7C(R)]$&            
\vr\cr\hrule
\+ 
\vrq n \quad & \vrq \qquad $-H$     
 \quad\qquad\
& \vrq $4\tr[S_8C(R)]$&            
\vr\cr\hrule
\+ 
\vrq p \quad & \vrq \qquad $-J$     
 \quad\qquad\
& \vrq $8\tr[S_5P]$&            
\vr\cr\hrule
\+ 
\vrq q \quad & \vrq \qquad $-J$
 \quad\qquad\
& \vrq $4\tr[S_5P]$&
\vr\cr\hrule
\+ 
\vrq r \quad & \vrq \qquad $-F$ 
 \quad\qquad\
& \vrq $4\tr[P^3C(R)]$&          
\vr\cr\hrule
\+
\vrq s \quad & \vrq \qquad $-F$
 \quad\qquad\
& \vrq $4\tr[P^3C(R)]$&
\vr\cr\hrule

}$$
\in
{\it \noindent
Table 2: Four-loop contributions to $<VV>_{\rm pole}$ for the 
diagrams in Figure 2. Each contribution is obtained by multiplying the 
simple pole from the momentum integral in the first column by the group
theory factor in the second column, and by $r^{-1}g^2$. Note that 
only contributions including terms of $O(Y^6)$ have been retained in the group 
theory factors.}
\out
\vfill\eject

\epsfysize= 7.5in
\centerline{\epsfbox{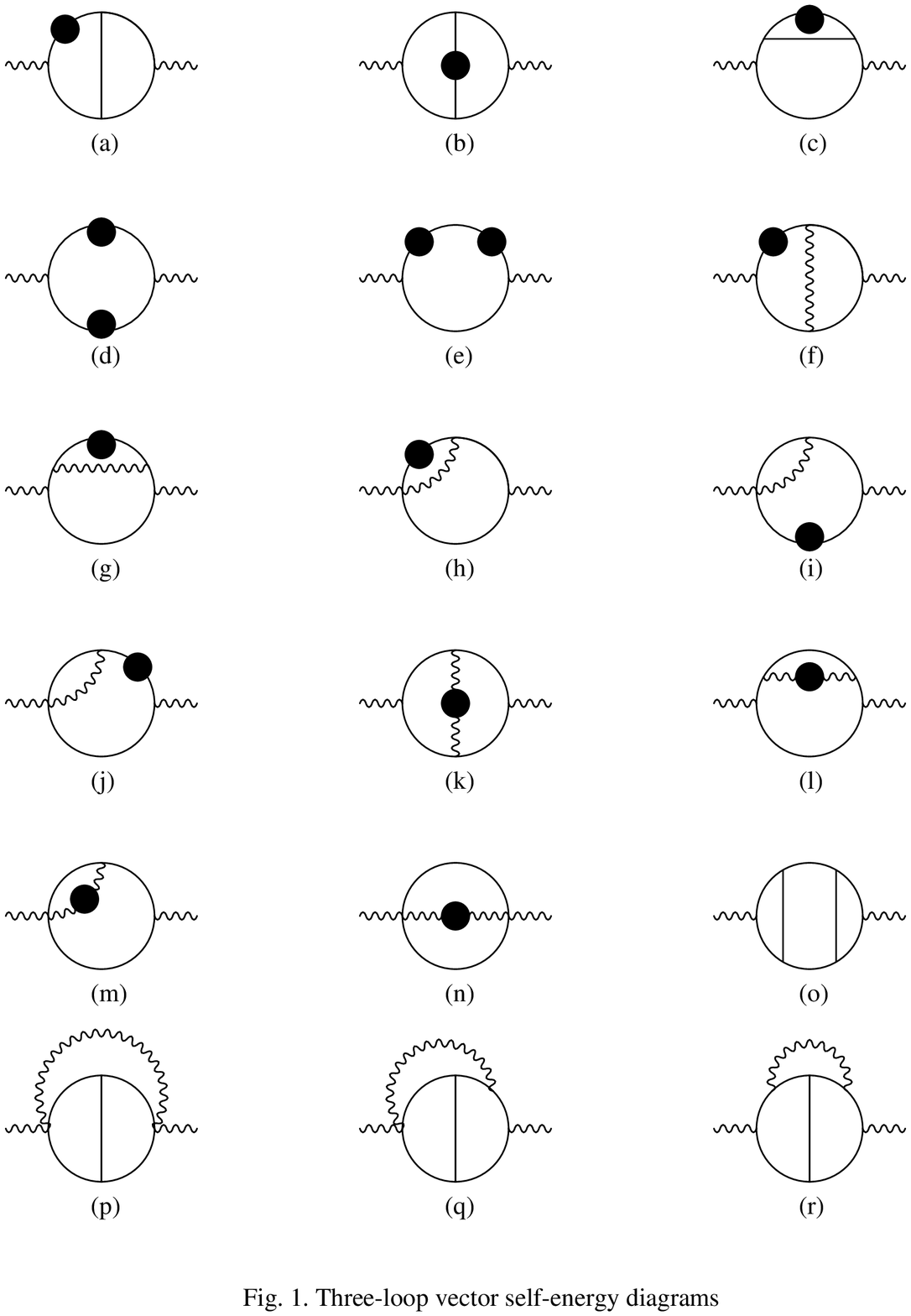}}

\epsfysize= 7.5in
\centerline{\epsfbox{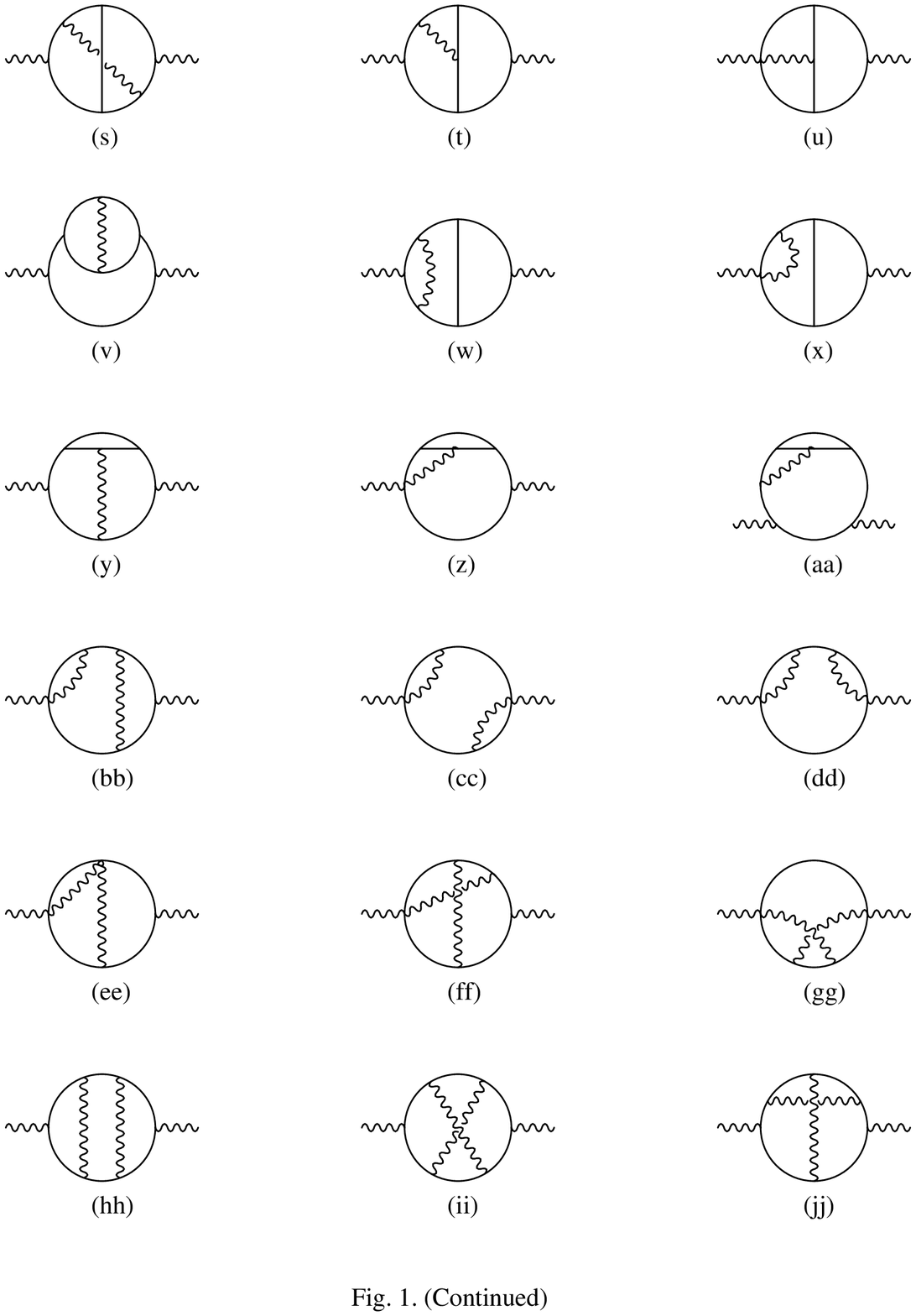}}

\epsfysize= 6.5in
\centerline{\epsfbox{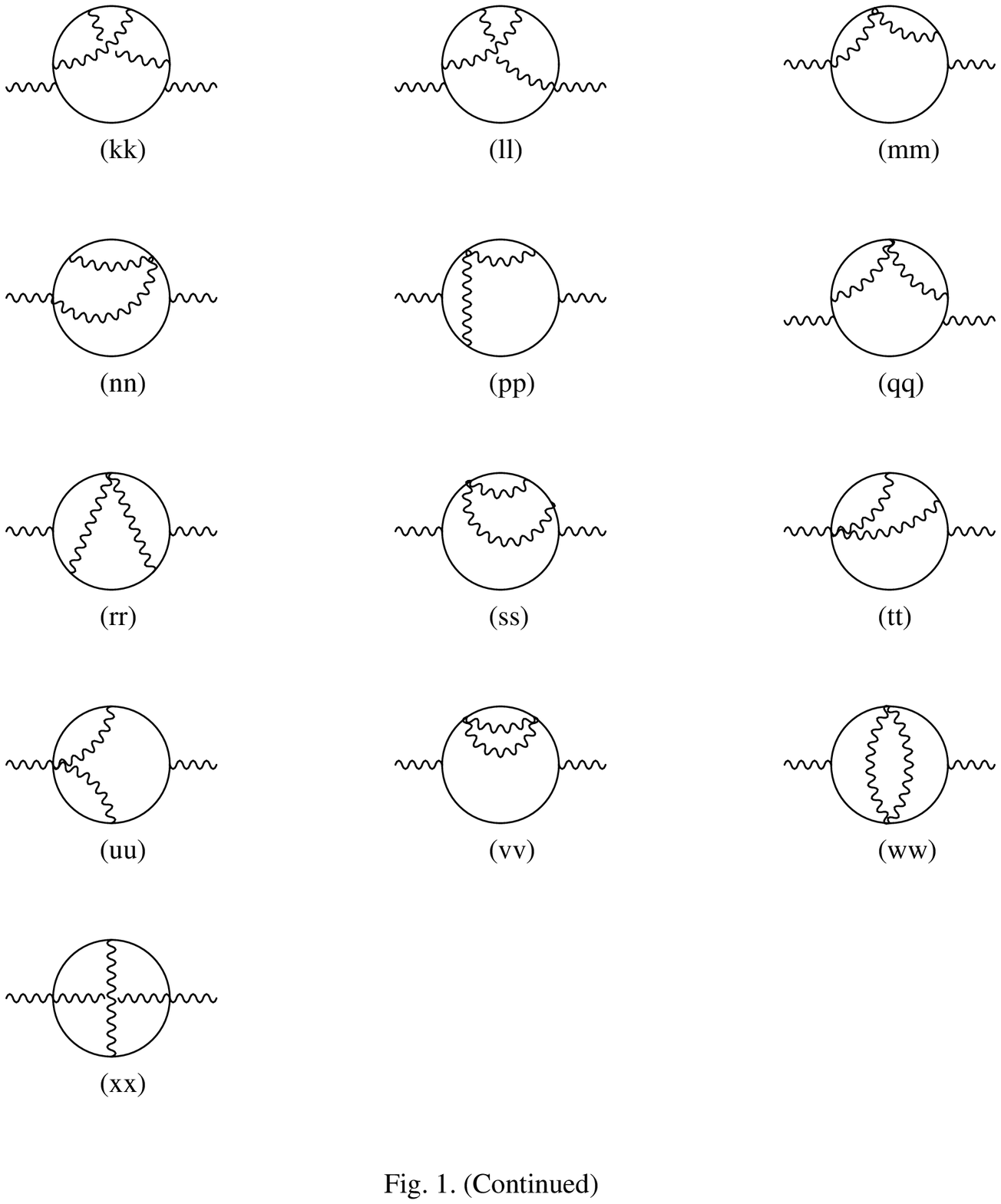}}
\vfill\eject
\epsfysize= 3in
\centerline{\epsfbox{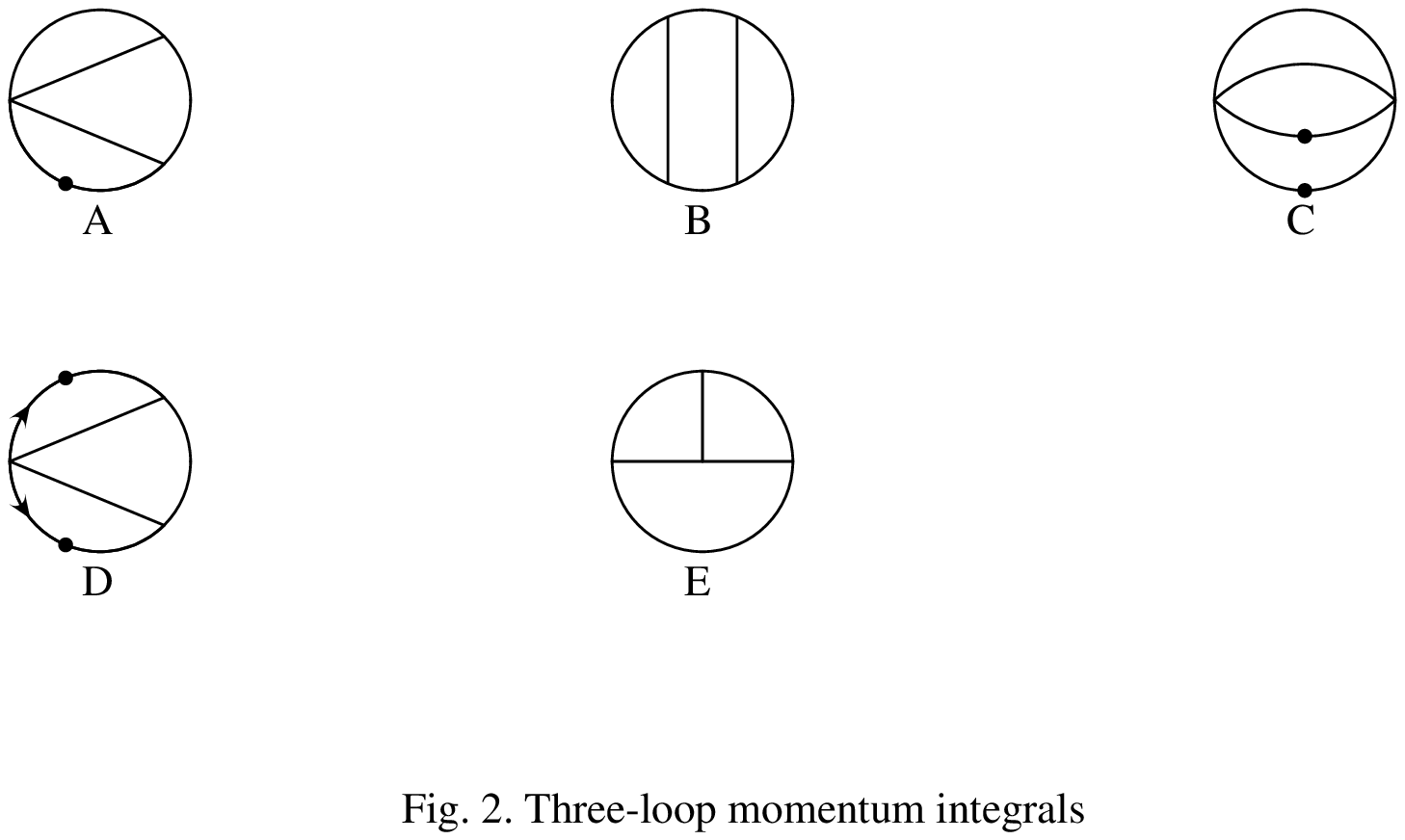}}

\epsfysize= 6.5in
\centerline{\epsfbox{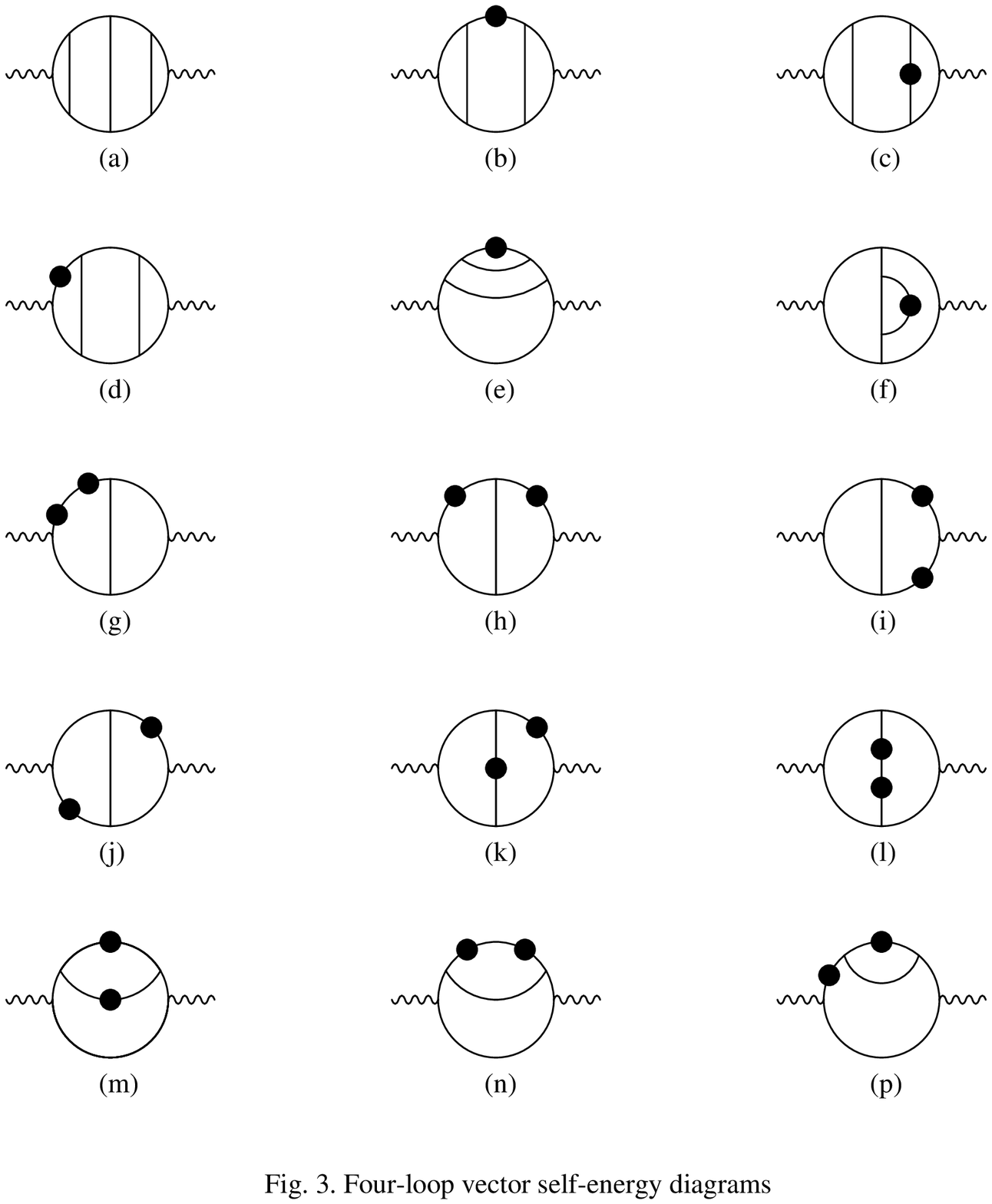}}

\epsfysize= 2.8in
\centerline{\epsfbox{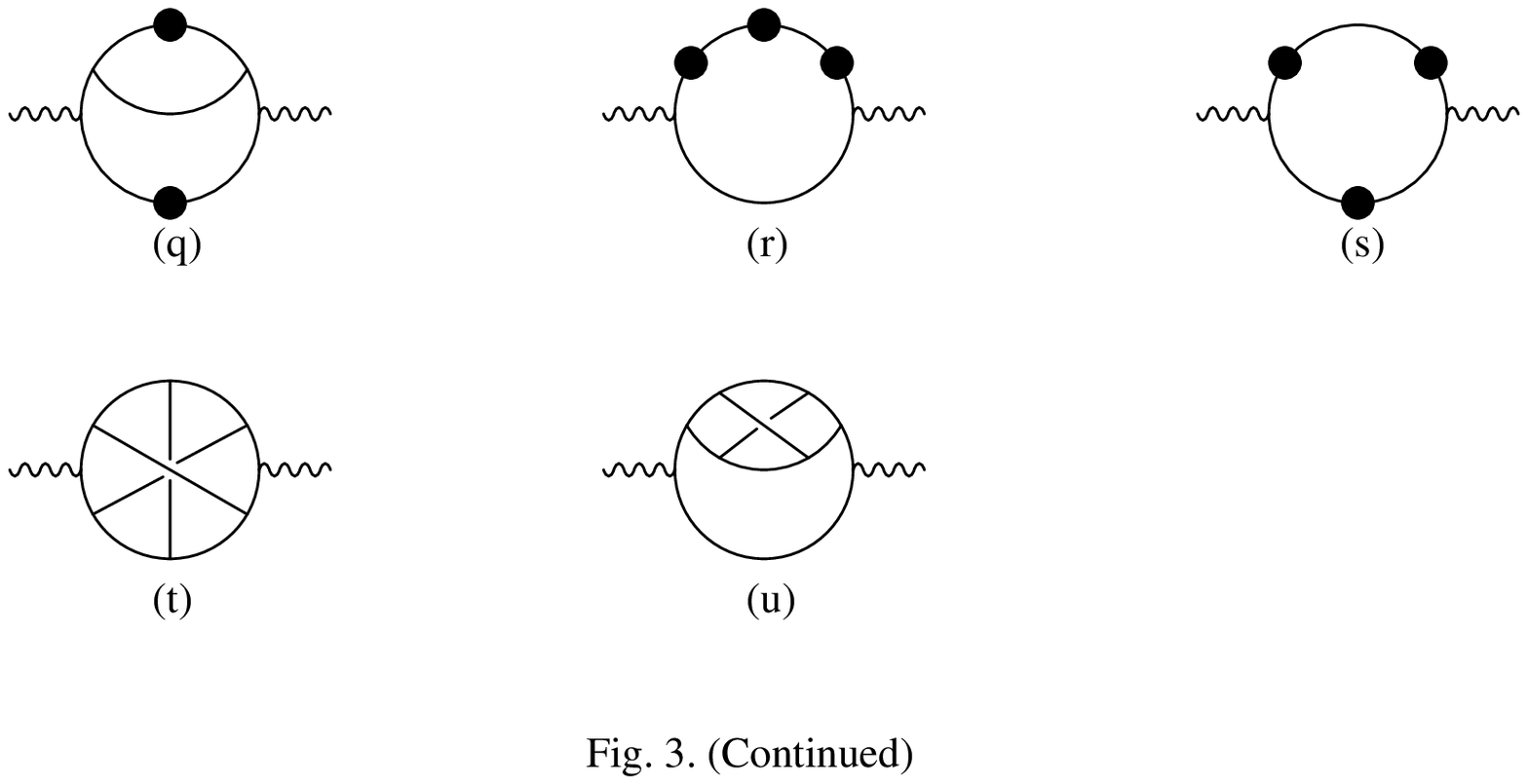}}
\vfill\eject
\epsfysize= 5.5in
\centerline{\epsfbox{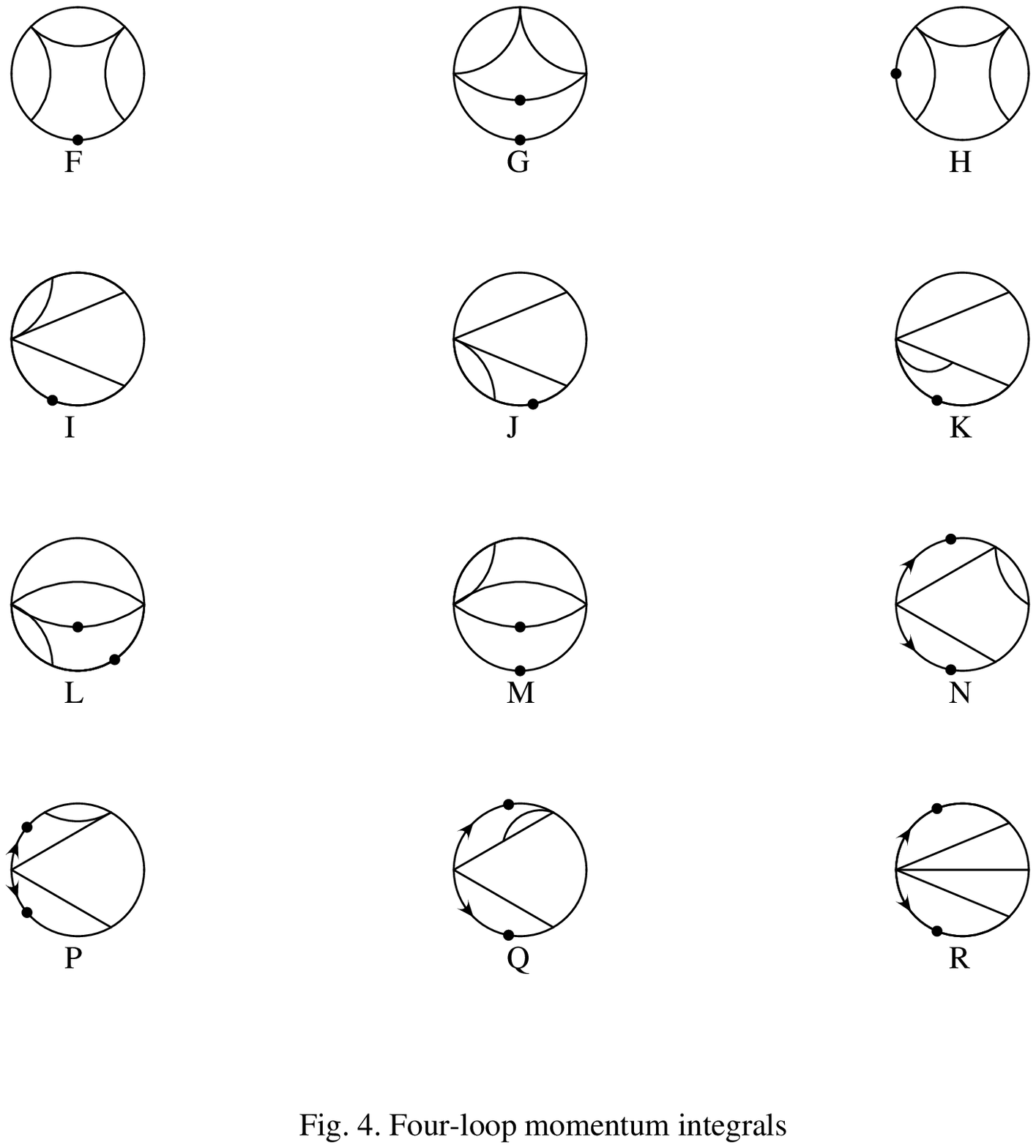}}

\bye